# LUMINESCENCE OF SPELEOTHEMS IN ITALIAN GYPSUM CAVES: PRELIMINARY REPORT

Yavor Y. Shopov[1], Diana Stoykova[1], Paolo Forti[2]

*Riassunto*

Utilizzando una lampada ad impulsi di Xenon si sono ottenuti gli spettri di luminescenza di tre campioni di concrezioni carbonatiche del sistema carsico dell'Acquafredda e di uno della Grotta Novella (Parco dei Gessi Bolognesi, Italia). Si ritiene che tutte queste concrezioni carbonatiche si siano formate utilizzando esclusivamente la $CO_2$ dell'atmosfera, dato che la roccia in cui si è sviluppata la grotta (i Gessi messiniani) non contiene carbonati. Le immagini ottenute rappresentano praticamente la registrazione delle variazioni climatiche verificatesi durante lo sviluppo degli speleotemi. Le datazioni U/Th e $^{14}C$ hanno evidenziato come le concrezioni abbiano iniziato a crescere circa 5000 anni fa. Lo studio di dettaglio degli spettri di luminescenza è ancora in corso.

**Parole chiave:** Luminescenza, Concrezionamento, Carsismo in gesso, Variazioni climatiche, Bologna, Italia.

*Abstract*

*The luminescence of 3 speleothem samples from the Acquafredda karst system and 1 from the Novella Cave (Gessi Bolognesi Natural Park, Italy) has been recorded using excitation by impulse Xe- lamp. All these carbonate speleothems are believed to be formed only from active $CO_2$ from the air, because the bedrock of the cave consist of gypsum and does not contain carbonates. The obtained photos of luminescence record the climate changes during the speleothem growth. U/Th and $^{14}C$ dating proved that studied speleothems started to grow since about 5,000 years ago. The detailed analyses of the luminescence records is still in progress.*

## Introduction

Calcite speleothems frequently display luminescence that is produced by calcium salts of humic and fulvic acids derived from soils above the cave (SHOPOV, 1989a, 1989b; WHITE & BRENNAN, 1989). These acids are released by the roots of living plants, and by the decomposition of dead vegetative matter. Root release is modulated by visible solar radiation via photosynthesis, while rates of decomposition depend exponentially on soil temperature. The soil temperature depends mainly on solar infrared and visible radiation (SHOPOV *et al.*, 1994) in case the cave is covered only by grass, or on air temperatures in case the cave is covered by forest or bush. In the first case, microzonality of luminescence

1 University Center for Space Research, Faculty of Physics, University of Sofia, James Baucher 5, Sofia 1164, Bulgaria. E-mail: YYShopov@Phys.Uni-Sofia.BG
2 Dipartimento di Scienze della Terra e Geologico Ambientali, Via Zamboni 67, 40127, Bologna, Italy.

of speleothems can be used as an indirect Solar Activity (SA) index (SHOPOV et al., 1990, 1991), but in the second case it can be used as a paleotemperature proxy (SHOPOV et al., 1996).

Luminescence organics in speleothems can be divided in four types:
1) Calcium salts of fulvic acids;
2) Calcium salts of humic acids;
3) Calcium salts of huminomelanic acids;
4) Organic esters.

All these four types are usually present in a single speleothem with hundreds of chemical compounds with similar chemical behaviour, but of different molecular weights. Concentration distribution of these compounds (and their luminescence spectra) depends on type of soils and plants over the cave, so the study of luminescent spectra of these organic compounds can give information about paleosoils and plants in the past (WHITE & BRENNAN, 1989). Changes in visible colour of luminescence of speleothems suggesting major changes of plant society are rarely observed.

Most known luminescent centres in calcite are inorganic ions of Mn, Tb, Er, Dy, U, Eu, Sm and Ce (SHOPOV, 1997). Minerals contain many admixtures. Usually several centres activate luminescence of the sample and the measured spectrum is a sum of the spectra of two or more of them. Luminescence of minerals formed at normal cave temperatures is normally due mainly to molecular ions and sorbed organic molecules. But luminescence of uranil-ion is also very common in such speleothems.

Experimental

The visible luminescence of two calcite speleothems from the Acquafredda karst system and of one calcite speleothem from Novella Cave (Gessi Bolognesi Natural Park) has been analysed. All the samples proved to be potentially really interesting for high resolution luminescence analyses.

Photos on fig.1-4 are negative photographs of integral phosphorescence of 2-5mm thick polished sections of the speleothem samples under excitation by impulse Xe-lamp in the entire UV and visible spectra. So darker parts of the image correspond to brighter luminescence, which in turn indicates an higher concentration of humic and fulvic acids, and perhaps to warmer climate (SHOPOV, 1997).

The most interesting among the studied samples is the calcite stalagmite SP1 (Fig. 1) from the Spipola Cave (Acquafredda karst system). The U/Th and $^{14}$C datings of the first layers of carbonate speleothems within the cave (but not of this sample) proved that all the speleothems in this karst system started to grow since about 5 thousand years ago (FORTI, 2003). Moreover all the carbonate speleothems from the gypsum karst of

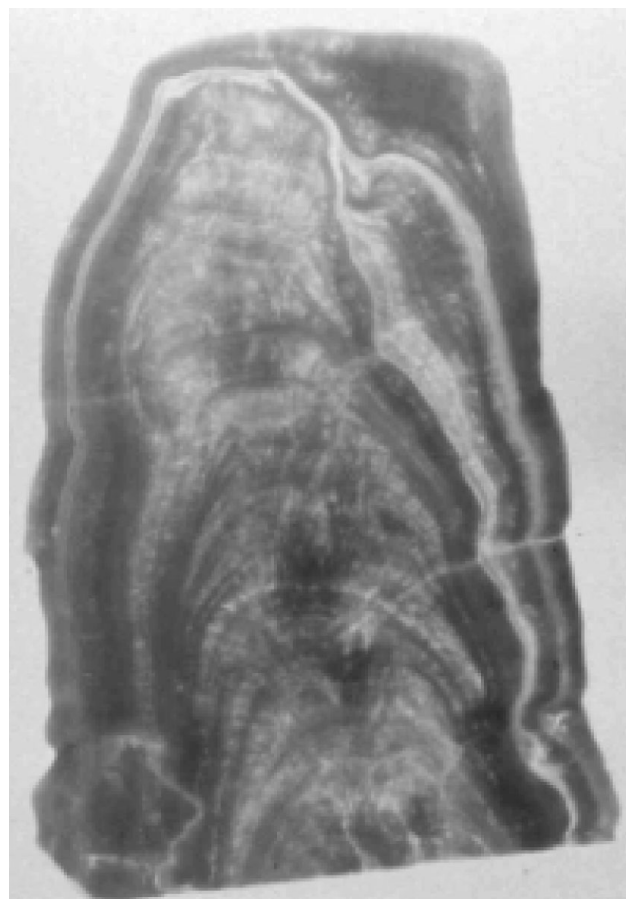

Fig. 1 - Luminescence of sample SP1 from the Spipola Cave (Acquafredda karst system).

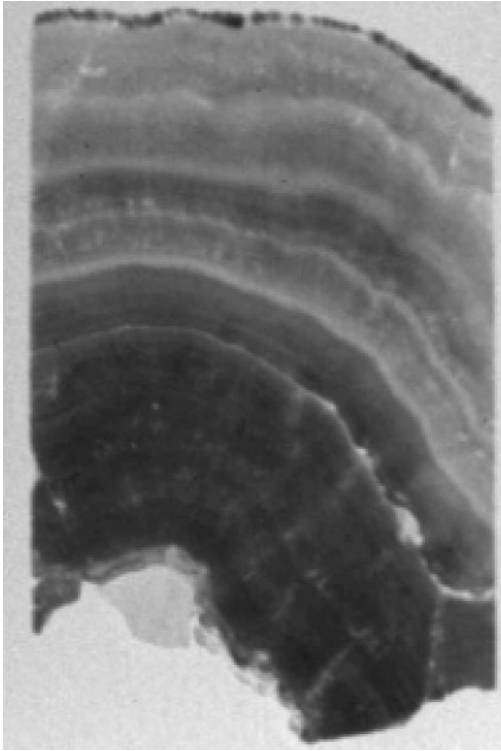

Fig. 2 - Luminescence of the sample NO1 from the Novella Cave.

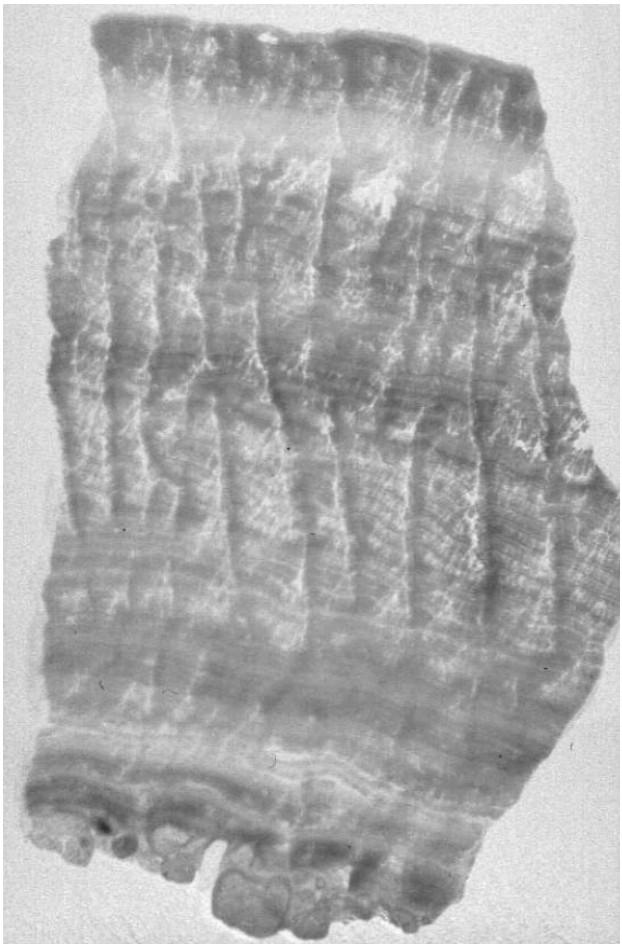

Fig. 3 - Luminescence of sample SP2 from the Spipola Cave (Acquafredda karst system).

Bologna are believed to be formed only from active $CO_2$ from the air, because the bedrock of the cave consist of gypsum and does not contain any significant amount of carbonate (FORTI & RABBI, 1981). Luminescence of this sample is very strong and exhibits many luminescence bands, which in turn make evident many variations in the direction of the growth axis of the speleothem, therefore this speleothem has been utilised for detailed reconstruction of earthquakes of the past (FORTI & POSTPISCHL, 1986). It was taken from the cave about 50 years ago (1950-1955). At that time the speleothem was active. The cave is covered only by thin layer (less than 50cm in average) of soil and a large portion of it is covered by an oak forest. The location of the main entrance is Lat. 44° 26' 47" / Long. 11° 22' 58".

A calcite flowstone from the Inner Doline in the Spipola Cave, which was some half meters thick has been analysed. U/Th and $^{14}C$ dating of it (FORTI, 2003) defined that its bottom is 5,000 and the top 2,000 years old. It has white colour with long fine crystals along the growth axes. Photos of luminescence of two pieces of it (SP2 and SP3) are presented on Fig. 3 and 4. Luminescence of the SP2 sample (Fig. 3) is strong and exhibits many fine luminescence bands. Because the half meters thick flowstone formed in 3,000 years or less, these bands seems to be annual. Luminescence of the SP3 sample (Fig. 4) is stronger but exhibits few luminescence bands. Finally a calcite flowstone (NO1) from Novella Cave, 3.5km far from Spipola (Fig. 2) was analysed: it was sampled about 30 years ago when it was still active. Luminescence of this sample is strong and exhibits several thick luminescence bands with relatively stable intensity of luminescence. They are separated by hiatuses. This suggests that this flowstone was growing relatively fast, but only during short periods of time (with stable climate). Such growth pattern suggests that the speleothem growth was possible only in a very small range of climatic conditions, so it did not grow most of the time. The location of the main entrance of the Novella Cave is at Lat. 44° 25' 38" / Long. 11° 24' 54". The situation of this cave is similar to that of Spipola: completely covered by oak forest.

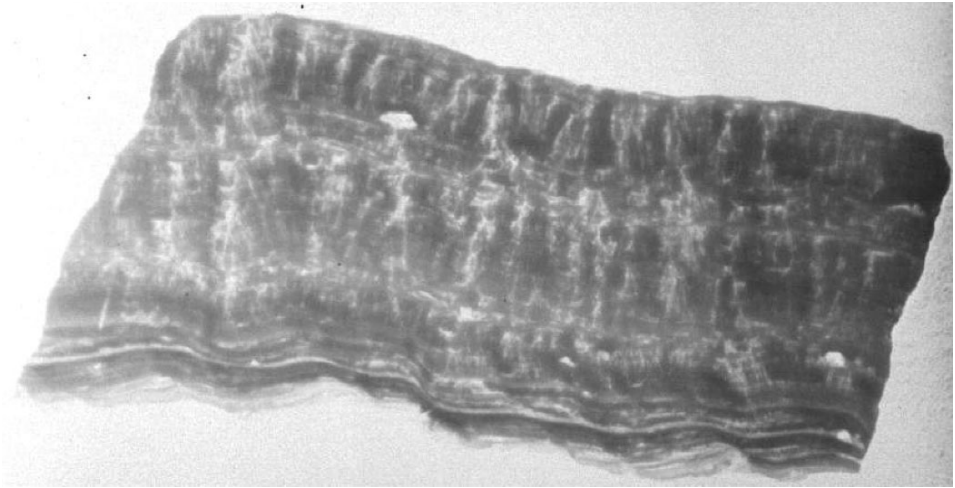

Fig. 4 - Luminescence of sample SP3 from the Spipola Cave (Acquafredda karst system).

The average temperature inside the two caves is about 10-11°C. Both caves are developed in Messinian gypsum. Further dating of the studied samples is necessary in order to convert obtained luminescence images into paleoclimatic records.

*Acknowledgements*
*The research has been made under grant NZ811 of Bulgarian Scientific Foundation to Y. Shopov and funded by project MURST-Cofin 2000, responsible Prof. Ugo Sauro, University of Padova.*